\documentclass{article}[12pt]

\font\sqi=cmssq8

\def\DR{\rm I\kern-1.45pt\rm R}

\def\DC{\kern2pt {\hbox{\sqi I}}\kern-4.2pt\rm C}

\newcommand{\ben}{\begin{enumerate}}

\newcommand{\een}{\end{enumerate}}

\newcommand{\beq}{\begin{equation}}

\newcommand{\eeq}{\end{equation}}

\newcommand{\bse}{\begin{subequation}}

\newcommand{\ese}{\end{subequation}}

\newcommand{\bea}{\begin{eqnarray}}

\newcommand{\eea}{\end{eqnarray}}

\newcommand{\bc}{\begin{center}}

\newcommand{\ec}{\end{center}}

\def\DH{\rm I\kern-1.5pt\rm H\kern-1.5pt\rm I}

\begin{document}
\begin{center}
{\large Noncommutative Quantum Scattering in a Central Field}\\
\vspace{0.5 cm} { \large Stefano Bellucci$^1$ and Armen
Yeranyan$^2$ }
\vspace{0.5 cm}

{\it $^1$ INFN-Laboratori Nazionali di Frascati,
 P.O. Box 13, I-00044, Frascati, Italy\\
$^2$ Yerevan State University, Alex  Manoogian St., 1, Yerevan,
375025, Armenia}
\end{center}
\vspace{0.4cm}

\begin{abstract}
In this paper the problem of noncommutative elastic scattering in
a central field is considered. General formulas for the differential
cross-section for two cases are obtained. For the case of high
energy of an incident wave it is shown that the differential
cross-section coincides with that on the commutative space. For
the case in which noncommutativity yields only a small correction to the
central potential it is shown that the noncommutativity leads to
the redistribution of particles along the azimuthal angle, although
the whole cross-section coincides with the commutative case.
\end{abstract}

\begin{center}


\end{center}

\setcounter{equation}0

\section{Introduction}

 In recent years considerable attention was paid to the
investigation of noncommutative spaces. The reasons for the
emergence of this interest were the predictions of String theory
\cite{string} in the low-energy limit which, along with the
Brane-world scenario \cite{bran}, led to the fact that the
space-time could be noncommutative. Later intensive investigation
of the field theory on noncommutative spaces was prompted by
M-theory \cite{cdsh} and the matrix formulation of the quantum Hall effect
\cite{hall}. Let us mention the fact that noncommutative spaces
may arise as the gravitation quantum effect and may serve as a
possible way for the regularization of quantum field
theories \cite{gqe}.

Noncommutative spaces are characterized by the fact that their
coordinate operators satisfy the equation
\begin{equation}
[x^i,x^j]=\imath \hbar \Theta^{i j},
\end{equation}
where $\Theta^{i j}$ is the constant parameter of
noncommutativity. The parameter $\Theta^{i j}$ is real, antisymmetric and
has the dimension  $\frac{(length)^2}{\hbar}$. In order to determine the
physical system on the noncommutative space (according to
\cite{moyal}), the Lagrangian of the corresponding commutative system
is taken and all the usual derivatives in it are replaced with the
(Weyl-Moyal) star product
\begin{equation}
(f \star g)(x)=e^{ \frac{\imath}{2}\hbar
\Theta^{ij}\partial_{x^i}\partial_{x'^j}}f(x)g({x'})|_{{x}={x'}},
\end{equation}
where $f$ and $g$ are infinitely differentiable functions.

The noncommutative deformation of the Standard model \cite{standart}
was suggested for the study of the phenomenological consequences
of noncommutativity.\footnote{
For the renormalization of the energy-momentum tensor in noncommutative field theories, see
\cite{bbk}).}
For these purposes quantum mechanics on
noncommutative spaces was also extensively studied. The following
steps were carried out: oscillator models on various noncommutative spaces were
constructed, hydrogen-like atoms on these spaces were thoroughly
considered. Stark, Zeeman and other effects were examined on
these spaces. See details in \cite{nqm} and references therein.

At the same time very little attention was payed to the study of the
quantum mechanical scattering problem on noncommutative spaces.
There is only one paper \cite{scat} where scattering on the
noncommutative plane is investigated. It is obvious that the
scattering problem is a key one because it may connect
noncommutative effects with the reality on the experimental level. The
consideration of the scattering problem is focal also for another
reason. As it was shown in \cite{kao}, in most "attraction" problems
noncommutativity does not play any role.

The purpose of the present work is to investigate the scattering in
the central field on the noncommutative space. In Section 2 the
two-body problems are briefly considered. In Section 3.1 the
scattering problem in the arbitrary central field for the case of
high energies (the Born approximation) is investigated. In Section
3.2 we study the case when the potential cannot be considered small and
another approximation has to be applied. In the
Conclusion the main results are summarized.

\setcounter{equation}{0}

\section{The Two Body Problem}

Let us consider a system of two quantum particles with respective
masses and charges $(m_1,\, e_1)$ and $(m_2,\, e_2)$ on a
noncommutative space. As it was shown in \cite{kao}, the
noncommutativity of a particle is proved to differ from its
antiparticle by the sign. Consequently the Moyal product for this
case can be written as

\begin{equation}
f(\tilde{\bf{r}}_1,\tilde{\bf{r}}_2)\star_{Z_1Z_2}\,g(\tilde{\bf{r}}_1,\tilde{\bf{r}}_2)=e^{
\frac{\imath}{2}\hbar\Theta^{ij}(Z_1\frac{\partial}{\partial\tilde{x}^i_1}
\frac{\partial}{\partial\tilde{x}'^j_1}+Z_2\frac{\partial}{\partial\tilde{x}^i_2}
\frac{\partial}{\partial\tilde{x}'^j_2})}f(\tilde{\bf{r}}_1,\tilde{\bf{r}}_2)
g(\tilde{\bf{r}}'_1,\tilde{\bf{r}}'_2)|_{\tilde{\bf{r}}=\tilde{\bf{r}}'},
\label{moyal}
\end{equation}
where $Z_k$ are charge numbers ($e_k=Z_k e$, $k=1,2$). This means
that the commutation relations for $\tilde{x}^i_k$ and momentum
operators $\hat{\tilde{p}}_{i(k)}$ take the form

\begin{equation}
[\tilde{x}^i_k,\tilde{x}^j_l]=\imath\hbar\Theta^{ij}Z_k\delta_{kl},\quad
[\tilde{x}^i_k,\hat{\tilde{p}}_{j(l)}]=\imath\hbar\delta^i_j\delta_{kl}.\label{comut}
\end{equation}

The two body problem Hamiltonian can be written as

\begin{equation}
\hat{H}=\frac{\hat{\tilde{\bf{p}}}^2_1}{2m_1}+\frac{\hat{\tilde{\bf{p}}}^2_2}{2m_2}+
V(|\tilde{\bf{r}}_1-\tilde{\bf{r}}_2|).
\end{equation}

For a further separation of variables let us pass to the "center
mass" system. For this purpose we introduce "relative" coordinates

\begin{equation}
\tilde{\bf{r}}=\tilde{\bf{r}}_1-\tilde{\bf{r}}_2
\end{equation}
and "center of mass" coordinates

\begin{equation}
\tilde{\bf{R}}=\frac{m_1\tilde{\bf{r}}_1+m_2\tilde{\bf{r}}_2}{m_1+m_2}.
\end{equation}

The appropriate momenta have the form

\begin{equation}
\hat{\tilde{\bf{p}}}=\frac{m_2\hat{\tilde{\bf{p}}}_1-m_1\hat{\tilde{\bf{p}}}_2}{m_1+m_2}
\end{equation}
and

\begin{equation}
\hat{\tilde{\bf{K}}}=\hat{\tilde{\bf{p}}}_1+\hat{\tilde{\bf{p}}}_2
\end{equation}
respectively. The corresponding Hamiltonian reads

\begin{equation}
\hat{H}=\frac{\hat{\tilde{\bf{p}}}^2}{2\mu}+\frac{\hat{\tilde{\bf{K}}}^2}{2M}+
V(|\tilde{\bf{r}}|) \label{ham}
\end{equation}
and the commutation relations are

\begin{eqnarray}
[\tilde{x}^i,\tilde{x}^j]=\imath\hbar\Theta^{ij}(Z_1+Z_2), \quad
[\tilde{x}^i,\tilde{X}^j]=\imath\hbar\Theta^{ij}\frac{m_1Z_1-m_2Z_2}{m_1+m_2}, \\
\nonumber
[\tilde{X}^i,\tilde{X}^j]=\imath\hbar\Theta^{ij}\frac{m_1^2Z_1+m_2^2Z_2}{(m_1+m_2)^2},
\quad
[\tilde{x}^i,\hat{\tilde{p}}_j]=[\tilde{X}^i,\hat{\tilde{K}}_j]=\imath\hbar\delta^i_j,
\end{eqnarray}
where $\mu$ and $M$ are the "reduced" and "total" masses.

One can also pass to the variables in which the commutation
relations have a canonical form. These new variables depend on the
previous ones in following way:

\begin{eqnarray}
{\bf{r}}=\tilde{\bf{r}}-\frac{1}{2}(Z_1+Z_2){\bf{\Theta}}\times{\hat{\tilde{\bf{p}}}}
&-&\frac{1}{2}\frac{m_1Z_1-m_2Z_2}{m_1+m_2}{\bf{\Theta}}\times{\hat{\tilde{\bf{K}}}},
\\ \nonumber
{\bf{R}}=\tilde{\bf{R}}-\frac{1}{2}\frac{m_1Z_1-m_2Z_2}{m_1+m_2}{\bf{\Theta}}\times{\hat{\tilde{\bf{p}}}}
&-&\frac{1}{2}\frac{m_1^2Z_1+m_2^2Z_2}{(m_1+m_2)^2}{\bf{\Theta}}\times{\hat{\tilde{\bf{K}}}},
\\ \nonumber
\hat{\bf{p}}=\hat{\tilde{\bf{p}}}, \quad
\hat{\bf{K}}&=&\hat{\tilde{\bf{K}}},
\end{eqnarray}
where $\Theta_{k}=\varepsilon_{kij}\Theta^{ij}$. Using them one
can write down the Hamiltonian (\ref{ham}) as follows:

\begin{equation}
\hat{H}=\frac{\hat{{\bf{p}}}^2}{2\mu}+\frac{\hat{{\bf{K}}}^2}{2M}+
V(|{\bf{r}}+\frac{1}{2}(Z_1+Z_2){\bf{\Theta}}\times{\hat{{\bf{p}}}}
+\frac{1}{2}\frac{m_1Z_1-m_2Z_2}{m_1+m_2}{\bf{\Theta}}\times{\hat{{\bf{K}}}}|).
\label{ham1}
\end{equation}

As it is seen, the Hamiltonian (\ref{ham1}) does not depend on $\bf{R}$
and consequently the momenta $\hat{\bf{K}}$ are preserved.
Shifting the origin of the coordinate system and neglecting the
constant kinetic energy of the center of mass, one can bring the
Hamiltonian (\ref{ham1}) to the form

\begin{equation}
\hat{H}=\frac{\hat{{\bf{p}}}^2}{2\mu}+
V(|{\bf{r}}+\frac{1}{2}(Z_1+Z_2){\bf{\Theta}}\times{\hat{{\bf{p}}}}
|).
\end{equation}

Taking into consideration the fact that ${\bf{\Theta}}$ is small and
keeping in the Hamiltonian only the first order terms on $\Theta$, we
get

\begin{equation}
\hat{H}=\frac{\hat{{\bf{p}}}^2}{2\mu}+
V(r)-\frac{Z_1+Z_2}{2r}\frac{dV(r)}{dr}{\bf{\Theta}}\cdot{\hat{\bf
L}},\label{ham2}
\end{equation}
where $r=\sqrt{|\bf{r}|}$ and $\hat{\bf
L}=\bf{r}\times\hat{\bf{p}}$ is the angular momentum operator.

It has to be mentioned that in this case, besides the energy, we have
two conserved quantities - i.e. the angular momentum magnitude and its
projection in the ${\bf{\Theta}}$  direction. It can be easily
seen by calculating the time derivative of the angular momentum operator
\cite{anmom}

\begin{equation}
\frac{d\hat{\bf L}}{dt}=[\hat{\bf
L},\hat{H}]=-\frac{Z_1+Z_2}{2r}\frac{dV(r)}{dr}{\bf{\Theta}}\times{\hat{\bf
L}}.
\end{equation}

\setcounter{equation}{0}

\section{Noncommutative Quantum Scattering}
\subsection{Born Approximation}

Let us consider elastic quantum scattering on noncommutative three
dimensional space. Our purpose is to compute the differential
cross-section. The Schr\"{o}dinger equation according to the
Hamiltonian (\ref{ham2}) has the form

\begin{equation}
-\frac{\hbar^2}{2\mu}\Delta\Psi({\bf{r}})+U({\bf{r}})\Psi({\bf{r}})=E\Psi({\bf{r}}),
\label{sh}
\end{equation}
where we introduced the notation

\begin{equation}
U({\bf{r}})=V(r)-\frac{Z_1+Z_2}{2r}\frac{dV(r)}{dr}{\bf{\Theta}}\cdot{\hat{\bf
L}}. \label{pot}
\end{equation}

The exact solution of equation (\ref{sh}), which describes
scattering, can be obtained also from the following integral
equation:

\begin{equation}
\Psi^{(\pm)}_a({\bf{r}})=e^{\imath \bf{k}_a
\bf{r}}-\frac{\mu}{2\pi\hbar^2}\int\frac{\exp(\pm\imath k
|\bf{r}-\bf{r'}|)}{|{\bf{r}}-{\bf{r'}}|}U({\bf{r'}})\Psi^{(\pm)}_a({\bf{r'}})d^3{\bf{r'}},
\label{int}
\end{equation}
where $\bf{k}_a$ is the wave vector of an incident plane wave and $k$ is
its magnitude. At large distances the solution of equation
(\ref{int}) corresponding to the scattering has the form

\begin{equation}
\Psi^{(+)}_a({\bf{r}}) \approx
\phi_a({\bf{r}})+\frac{f^{(+)}_a(\theta,\varphi)}{r}e^{\imath k r},
\end{equation}
where the first term is an incoming plane wave and the second one is
an outgoing spherical wave. Here

\begin{equation}
f^{(+)}_a(\theta_b,\varphi_b)=-\frac{\mu}{2\pi\hbar^2}\langle
\phi_b|U|\Psi^{(+)}_a\rangle \label{ampl}
\end{equation}
denotes a scattering amplitude and $\phi_b$ is an outgoing plane wave.
Differently from the case of the central field it depends also on
the azimuthal angle. The differential cross-section $d\sigma$ is
expressed in terms of the scattering amplitude as follows:

\begin{equation}
d\sigma=f^{(+)}_a(\theta_b,\varphi_b)\sin\theta_b \, d \theta_b d
\varphi_b.
\end{equation}

The formula (\ref{ampl}) is exact. In order to find the scattering
amplitude we have to apply certain assumptions. In this Section we
will assume that the energy $E$ of the relative motion is large
enough, i.e. $V(r)<<E$ (as it was mentioned the two-dimensional
analog was investigated in \cite{scat}). Consequently,
$\Psi^{(+)}_a$ is very little different from the incoming plane wave
$\phi_a$ and in (\ref{ampl}) we can substitute it by the latter
one. This leads us to the first Born approximation

\begin{equation}
f^{(+)}_a(\theta_b,\varphi_b)=-\frac{\mu}{2\pi\hbar^2}\langle
\phi_b|U|\phi_a\rangle. \label{ampl1}
\end{equation}

As it is seen from formulae (\ref{pot}) and (\ref{ampl1}) the
scattering amplitude consists of two parts. The first one, which
depends only on  "scattering" angle, is an amplitude in the
central field on commutative space

\begin{equation}
f_{com}(\theta)=-\frac{\mu}{2\pi\hbar^2}\langle
\phi_b|V(r)|\phi_a\rangle, \label{com}
\end{equation}
whereas the second one is the addition due to noncommutativity

\begin{equation}
f_{noncom}(\theta,\varphi)=\frac{(Z_1+Z_2)\mu}{4\pi\hbar^2}\langle
\phi_b|\frac{1}{r}\frac{dV(r)}{dr}{\bf{\Theta}}\cdot{\hat{\bf
L}}|\phi_a\rangle. \label{noncom}
\end{equation}

In order to simplify formulae (\ref{com}), (\ref{noncom}) let us
introduce spherical coordinates, directing the $z$ axis along an incident
plane wave. Applying the plane wave expansion in
spherical harmonics presented below

\begin{equation}
\phi=e^{\imath \bf{k} \bf{r}}=4\pi
\sum^{\infty}_{l=0}\sum^{m=l}_{m=-l} \imath^l j_l(kr)Y^\ast_{l,\,
m}\left(\frac{\bf{k}}{k}\right)Y_{l,\,m}\left(\frac{\bf{r}}{r}\right),
\label{plos}
\end{equation}
where $j_l(kr)$ is a Bessel spherical function, $l$ and $m$ are orbital
and azimuthal quantum numbers  respectively, integrating over the angles
and considering the following equations:

\begin{equation}
\hat{L}_zY_{l,\,0}=0, \quad
\hat{L}_{\pm}Y_{l,\,0}=\hbar\sqrt{l(l+1)}\,Y_{l,\,\pm1}
\end{equation}
and sums

\begin{eqnarray}
\sum^{\infty}_{l=0}(2l+1)j^2_l(k r)P_l(\cos \theta)=j_0(q r), \\
\nonumber \sum^{\infty}_{l=1}(2l+1)j^2_l(k r)P^1_{l}(\cos
\theta)=k r \cos\frac{\theta}{2}\,j_1(q r),
\end{eqnarray}
where $q=|{\bf{k}}_a-{\bf{k}}_b|=2k\sin\frac{\theta}{2}$, $P_l$
and $P^1_{l}$ are usual and adjoined Legendre polynomials, we
obtain

\begin{eqnarray}
f_{com}=-\frac{2\mu}{\hbar^2}\int^{\infty}_0 V(r) j_0(q r) r^2 d
r, \label{c} \\
f_{noncom}=-\imath
({\bf{\Theta}}\cdot{\bf{n}})\frac{2\mu(Z_1+Z_2)}{\hbar}\frac{k^2}{q}\sin
\theta \int^{\infty}_0 \frac{d V(r)}{d r} j_1(q r) r^2 dr.
\label{n}
\end{eqnarray}

Here $n$ is a unit vector which is normal to the scattering plane and is
determined by the equation

\begin{equation}
{\bf{k}}_a\times {\bf{k}}_b={\bf{n}}\,k^2\sin\theta.
\end{equation}

Its components can be expressed by the azimuthal angle in the
following way:

\begin{equation}
{\bf{n}}=-\sin\varphi \, {\bf{i}}+\cos\varphi \, {\bf{j}},
\end{equation}
where ${\bf{i}}$ and {\bf{j}} are cartesian basis vectors. Using
the equation $x^2j_0(x)=\frac{d}{dx}(x^2j_1(x))$ and integrating
(\ref{c}) by parts, one can write

\begin{equation}
f=\frac{2\mu}{\hbar^2q}\left(1-\imath (Z_1+Z_2) \hbar k^2\sin
\theta ({\bf{\Theta}}\cdot{\bf{n}})\right)\int^{\infty}_0 \frac{d
V(r)}{d r} j_1(q r) r^2 dr. \label{f}
\end{equation}

It can be easily seen from the equation above that the
noncommutativity addition in the differential cross-section is of the
second order in $\Theta$. Taking into consideration that $\Theta$
is small enough one can say that, in the Born approximation, we have no
corrections due to noncommutativity.

\subsection{Disordered-Wave Born Approximation}

The Born approximation is used only when the potential is weak
enough to give very rapid convergence. In this Section we
investigate the case when the potential $V(r)$ is not small, and
for this we need an alternative approach. The disordered-wave Born
approximation (DWBA) can serve as such, since our potential
$U({\bf r})$ decomposes naturally into two parts: $U({\bf
r})=V(r)+W({\bf r})$, where $W({\bf
r})=-\frac{Z_1+Z_2}{2r}\frac{dV(r)}{dr}{\bf{\Theta}}\cdot{\hat{\bf
L}}$. The first term is the primary potential and the second one is a
small addition. It should be mentioned that this division is
especially useful if the scattering wave function under the action
of a primary part is obtained exactly. For example, as a primary
potential one can take the exactly soluble Coulomb potential.

Let $\chi^{(+)}_a$ be a scattering wave function of an unperturbed
potential (all notations are in accordance with the notations
of the previous Section), then the exact expression of a
scattering amplitude has the following form:

\begin{equation}
f=-\frac{\mu}{2\pi\hbar^2}\langle
\phi_b|V(r)|\chi^{(+)}_a\rangle-\frac{\mu}{2\pi\hbar^2}\langle
\chi^{(-)}_b|W({\bf{r}})|\Psi^{(+)}_a\rangle, \label{ampf}
\end{equation}
where $\Psi^{(+)}_a$ is a scattering wave function of a perturbed
Hamiltonian. The first term is the scattering amplitude in the
absence of a perturbing potential and the second one is a
correction due to $W({\bf{r}})$. Supposing $W({\bf{r}})$
to be small enough, one can change $\Psi^{(+)}_a$ for $\chi^{(+)}_a$
in the (\ref{ampf}). Then, for the above-mentioned correction we
have

\begin{equation}
\langle \chi^{(-)}_b|W({\bf{r}})|\chi^{(+)}_a\rangle=\int
\chi^{(-)\ast}_b({\bf{r}}) W({\bf{r}}) \chi^{(+)}_a({\bf{r}})
d^3{\bf{r}},\label{cor}
\end{equation}
with the first-order exactness by $W({\bf{r}})$. We can simplify
this formula by passing to spherical coordinates as it was done in
the previous Section.  The only difference is that in the latter
case we must use the expansion of $\chi^{(+)}_a$ on spherical
harmonics instead of (\ref{plos}). This expansion has the
following form:

\begin{equation}
\chi^{(\pm)}_{\bf
k}=\frac{4\pi}{kr}\sum^{\infty}_{l=0}\sum^{m=l}_{m=-l} \imath^l
e^{\pm \imath \eta_l} F_l(k;r)Y^\ast_{l,\,
m}\left(\frac{\bf{k}}{k}\right)Y_{l,\,m}\left(\frac{\bf{r}}{r}\right),\label{wav}
\end{equation}
where $ F_l(k;r)$ is a regular solution (corresponding to
scattering) of an unperturbed radial Schr\"{o}dinger equation. The coefficient
$\eta_l$ is a phase shift of the $l$-wave. Substituting (\ref{wav})
into (\ref{cor}) we get

\begin{equation}
\langle
\chi^{(-)}_b|\frac{1}{2r}\frac{dV(r)}{dr}{\bf{\Theta}}\cdot{\hat{\bf
L}}|\chi^{(+)}_a\rangle=\imath({\bf{\Theta}}\cdot{\bf{n}})
\frac{2\pi\hbar}{k^2}\sum^{\infty}_{l=1}(2l+1)P^1_{l}(\cos
\theta)e^{2\imath \eta_l
}\int^{\infty}_0\frac{1}{r}\frac{dV(r)}{dr}|F_l(k;r)|^2dr.
\end{equation}

As seen from the above formula, the noncommutative correction
permits to consider the scattering of each partial wave
separately. This correction is not present for the s-wave. As
distinguished from the case of the Born approximation, in this case
we have a first-order correction in $\theta$. The
dependence of this correction on the azimuthal angle is the same
for all partial waves. It leads to such a redistribution of
scattering particles by the angle $\varphi$, that integrating over this
angle results in the disappearance of this correction, i.e. to
the first order in $\theta$ we will have the same scattering
amplitude for the complete scattering cone, as in the case of the
unperturbed potential.

\section{Conclusion}

In this paper the problem of elastic noncommutative scattering in
an arbitrary central field was considered. Two cases were
investigated: firstly, the case when the whole potential can be considered
small and secondly, the one when the noncommutativity is considered as a
small addition to the main potential. In the first case it was
established that, to the first order in the noncommutativity parameter, there
are no additions to the formula of the "usual" scattering in an
arbitrary central field. In the second case we came to the
following results. It is estimated that the scattering of each
partial wave as in the usual case of central scattering
(commutative) can be considered separately. The scattering is
independent from the azimuthal quantum number and the dependence on
the azimuthal angle is the same for all types of waves.
The s-wave differential cross-section is like in the case when the
noncommutativity is absent. Differently with respect to the case with
low potential, there is a first-order addition to the differential
cross-section. The characteristic feature of this correction is
that the scattering in the full scattering cone is like in the case
when noncommutativity is absent, although it leads to the
redistribution of particles along the azimuthal angle.

\subsection*{{ Acknowledgments}}

We express our gratitude to Roland Avagyan and Levon Mardoyan
for helpful discussions. Special thanks go to Armen Nersessian for
useful discussions and collaboration at the early stage of this work.
This research was partially supported by the European
Community's Marie Curie Research Training Network under contract
MRTN-CT-2004-005104 Forces Universe, as well as by INTAS-00-00254 grant.

\end{document}